\documentclass[aps,twocolumn,superscriptaddress,showpacs,amssymb,prb,floatfix,preprintnumbers]{revtex4-1}
\usepackage{graphicx,color}
\usepackage{dcolumn}
\usepackage{amsmath}
\usepackage{amssymb}

\usepackage[dvipdfm]{hyperref}
\hypersetup{colorlinks=true,linkcolor=blue,pagebackref=true,
  implicit=true,breaklinks=true,pagebackref=true,backref=true,
  bookmarks=true,bookmarksnumbered=true,hyperfootnotes=true,debug=true,
  naturalnames=false,citecolor=blue,pdfview=FitH,pdfstartview=FitH,hyperindex=true}

\newcommand{\urusi}     {URu$_2$Si$_2$}

\begin{document}

\thispagestyle{myheadings}

\title{Nuclear magnetic resonance studies of pseudospin fluctuations in URu$_2$Si$_2$}

\author{K. R. Shirer}
\affiliation{Department of Physics, University of California, Davis, CA 95616, USA}
\author{J. T. Haraldsen}
\affiliation{Theoretical Division, Los Alamos National Laboratory, Los Alamos, NM 87545, USA}
\affiliation{Center for Integrated Nanotechnologies, Los Alamos National Laboratory, Los Alamos, NM 87545, USA}
\author{A. P. Dioguardi}
\author{J. Crocker}
\author{N. apRoberts-Warren}
\author{A. C. Shockley}
\author{C.-H. Lin}
\author{D. M. Nisson}
\affiliation{Department of Physics, University of California, Davis, CA 95616, USA}
\author{J. C. Cooley}
\affiliation{Los Alamos National Laboratory, Los Alamos, New Mexico 87545, USA}
\author{M. Janoschek}
\thanks{Current Address: Condensed Matter and Magnet Science Group, Los Alamos National Laboratory, Los Alamos, NM 87545, USA}
\affiliation{Department of Physics, University of California, San Diego, La Jolla, California 92093-0319, USA}
\author{K. Huang}
\author{N. Kanchanavatee}
\author{M. B.  Maple}
\affiliation{Department of Physics, University of California, San Diego, La Jolla, California 92093-0319, USA}
\author{M. J. Graf}
\affiliation{Theoretical Division, Los Alamos National Laboratory, Los Alamos, NM 87545, USA}
\author{A. V. Balatsky}
\affiliation{Theoretical Division, Los Alamos National Laboratory, Los Alamos, NM 87545, USA}
\affiliation{Center for Integrated Nanotechnologies, Los Alamos National Laboratory, Los Alamos, NM 87545, USA}
\affiliation{NORDITA, Roslagstullsbacken 23, 106 91 Stockholm, Sweden}
\author{N. J. Curro}
\email{curro@physics.ucdavis.edu}
\affiliation{Department of Physics, University of California, Davis, CA 95616, USA}

\date{\today}

\begin{abstract}

We report $^{29}$Si NMR measurements in single crystals and aligned powders of URu$_2$Si$_2$ in the hidden order and paramagnetic phases.   The spin-lattice-relaxation data reveal evidence of pseudospin fluctuations of U moments in the paramagnetic  phase.  We find evidence for partial suppression of the density of states below 30 K, and analyze the data in terms of a two component spin-fermion model.  We propose that this behavior is a realization of a
pseudogap between the hidden order transition $T_{HO}$ and 30 K. This behavior is then compared to other materials that demonstrate precursor fluctuations in a pseudogap regime above a ground state with long-range order.

\end{abstract}


\maketitle


Despite more than twenty years since its discovery, \urusi\ continues to attract considerable interest in the condensed matter community. \cite{PalstraURSdiscovery,MydoshReview}  This heavy fermion system develops a ``hidden" order phase below $T_{HO} =17.5$ K, and an unconventional superconducting state below 1.5 K. \cite{PalstraURSdiscovery,Kasahara2007}  The nature of the hidden order (HO) phase remains controversial, but it clearly does not involve magnetic ordering of dipole moments. \cite{BroholmURS} It may involve order of higher order multipoles, \cite{Hanzawa2007} exotic spin, orbital or spin-orbital density waves, \cite{MapleURu2Si2,chandraURS,zhitomirsky,Das2012,RiseburoughURS}
or hybridization between local moments and conduction electrons. \cite{Elgazzar2008,Chandra2013,DubiBalatskyPRL2011}
Extensive neutron scattering and angle-resolved photoemission work has suggested that it has an itinerant nature and involves some type of Fermi surface instability. \cite{BroholmURS,WiebeURS,FlouquetJPSL2010,Dakovski2011,Chatterjee2012,BoariuPRL2013}

Recent evidence has suggested that the hidden order is intimately connected with the onset of coherence of the Kondo lattice. \cite{hewson,HauleURSnature2009,DavisURSnature,YazdaniURS}  In Kondo lattice systems the 5$f$ electrons of the U are partially screened by the conduction electrons, leading to a renormalization of the electronic dispersion near the Fermi level.  At high temperatures the 5$f$ electrons remain localized and scatter the itinerant conduction electrons. \cite{hewson,HauleURSnature2009,DavisURSnature,YazdaniURS}  Below a coherence temperature, $T_{coh}$, the $f$ electrons hybridize with the conduction electrons, and the electronic dispersion reflects renormalized heavy quasiparticles. In other Kondo lattice systems coherence emerges as a crossover, but recent scanning tunneling microscopy (STM) results suggest that $T_{coh}$ coincides with $T_{HO}$, and thus the HO parameter is in fact the hybridization gap. \cite{DavisURSnature,YazdaniURS,DubiBalatskyPRL2011}

A re-examination and re-analysis of previous thermodynamic and neutron scattering measurements under the context of HO gap fluctuations have revealed the possible existence of a pseudogap occurring before the HO state and starting around 30 K. \cite{BalatskyURSPG} This analysis highlighted the presence of anomalies in magnetic susceptibility, \cite{MapleURu2Si2} point contact spectroscopy (PCS), \cite{Hasselbach1992} and neutron scattering measurements. \cite{WiebeURS,Janik2009} Other thermodynamic probes (heat capacity, thermal expansion and ultrasound velocity) also register an anomaly between 25-30 K, where these are sensitive to changes in the elastic constants of the crystal lattice. \cite{Schlabitz,vanDijkURS1997,deVisserURS1986,wolfJLTP1994} A similar temperature scale has been observed in ultrafast and conventional optical spectroscopy, which found a suppression of low energy spectral weight below 30 K that may be associated with a pseudogap. \cite{BonnURL1988,LiuPRB2011,OpticalHybridGapURS2011,OpticalHybridGapURS2011}

In contrast to the pseudogap and the aforementioned STM results, recent PCS work has attributed the suppression of conductance below 28 K to a hybridization gap. \cite{ParkPCSurs2012} As mentioned above, this suppression emerges well above $T_{HO}$ and has been observed in earlier reports occurring around 22 K. \cite{Hasselbach1992} In general, reports concerning a suppression of the density of states (DOS) above $T_{HO}$ have been mixed. Some specific heat and resistivity measurements \cite{MydoshReview} as well as several PCS experiments \cite{RodrigoURS1997,EscuderoURS1994,Naidyuk2001,Naidyuk1996157,Nowack1992,RonningURS2012} revealed no anomalies above $T_{HO}$. These discrepancies makes the overall results unreliable for the assessment of a pseudogap. We therefore look to investigate similar anomalies in nuclear magnetic resonance (NMR) measurements.

Here, we report new NMR spin-lattice-relaxation measurements on both single crystal and powder samples of URu$_2$Si$_2$ that indicate the presence of spin fluctuations that become partially gapped below a temperature $T_{pg}\approx 30$ K. These observations are consistent with the emergence of a pseudogap prior to the HO state. Using a pseudospin fluctuation model, we show that fluctuations alone cannot explain the suppression of the NMR response between $T_{HO}$ and 30 K. We therefore conclude that the existence of a pseudogap is needed to further suppress the modes of the hidden order.

Resistivity and NMR Knight shift measurements indicated that a third temperature scale, $T_{coh} \approx 80$ K, is also evident. \cite{ShirerPNAS2012} However, while the 30 K feature is seen in the four-point correlation function of the spin-lattice-relaxation rate, it is  not evident in the two-point correlation function of the Knight shift. This may be due to the difference between the dynamic and static scaling factors.



\begin{figure}
\includegraphics[width=\linewidth]{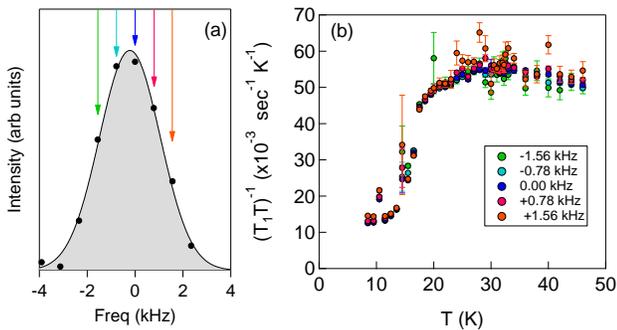}
\caption{(Color Online)
(a) The $^{29}$Si spectrum at 18 K indicating the width of the NMR resonant frequency, centered at 99.695 MHz, and the frequencies around at which $(T_1T)^{-1}$ was measured.
(b) $(T_1T)^{-1}$ as function of temperature for various values of frequency showing the consistency across the frequency range.
}
\label{fig:freq}
\end{figure}

Polycrystalline samples of URu$_2$Si$_2$ were synthesized by arc-melting in a gettered argon atmosphere, and an aligned powder was prepared in an epoxy matrix by curing a mixture of powder and epoxy in an external magnetic field of 9 T. Single crystals were grown employing the Czochralski method in a Techno Search TCA 4-5 Tetra-Arc Furnace under a zirconium gettered argon atmosphere. The grown single crystals were confirmed in a D-8 Discover Bruker diffractometer. In conducting samples, aligned powders are useful to enhance the surface-to-volume ratio and hence the NMR sensitivity, and single crystals were measured to confirm consistency between independent samples. NMR measurements were carried out using a high homogeneity 11.7 T (500 MHz) Oxford Instruments magnet. In this field, $T_{HO}$ is suppressed to 16 K. \cite{JaimeURSPRL2002}  The $^{29}$Si ($I=1/2$, natural abundance 4.6\%) spectra were measured by spin echoes, and the signal-to-noise ratio was enhanced by summing several ($\sim 100$) echoes acquired via a Carr-Purcell-Meiboom-Gill pulse sequence. The spin-lattice-relaxation rate $T_1^{-1}$ was measured as a function of the angle between the alignment axis and the magnetic field $\mathbf{H}_0$ in order to properly align the sample, since $T_1^{-1}$ is a strong function of orientation with a minimum for $\mathbf{H}_0 \parallel c$.

\begin{figure}
\includegraphics[width=\linewidth]{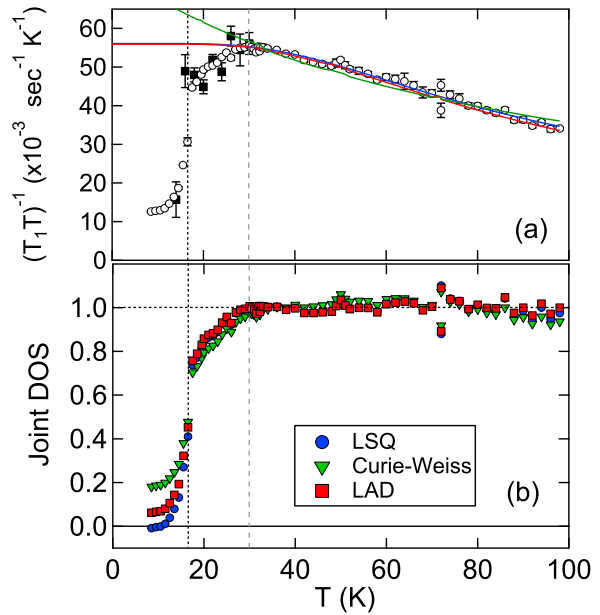}
\caption{(Color Online)
(a) Frequency-averaged $(T_1T)^{-1}$  versus $T$ for the aligned powder (open circles), and the single crystal (filled squares).  Different fits to 2D AFM spin fluctuations as described in Eq.~\ref{eqn:T1fit}  using an LAD  (red solid line)  and LSQ (blue solid line) fit method, as well as a LAD fit to a Curie-Weiss law (green solid line). (b) The extracted joint density of states (JDOS) showing the pseudogap suppression below 30 K and the residual carrier density in the HO state; same color legend as in panel (a). The JDOS drops by roughly 25\% in the pseudogap region between 30 K and 16 K and by another 70\% below the HO state. }
\label{fig:T1Tinv}
\end{figure}

Figure \ref{fig:freq} shows the measured NMR relaxation rate as function of frequency and temperature. There is no significant frequency dependence; we therefore use the average over all frequencies to determine the ($T_1T$)$^{-1}$, shown in Fig.~\ref{fig:T1Tinv}. Below $T_{HO}$, there is a clear change in the relaxation rate, as can be expected due to the development of long range order.  However, there is also a clear suppression of $(T_1T)^{-1}$ between 30 K and $T_{HO}$, which is indicative of the formation of a pseudogap. This linear pseudogap response is similar to the response predicted using gap fluctuations of the HO parameter by Haraldsen \textit{et~al.}. \cite{BalatskyURSPG}
Figure~\ref{fig:T1Tinv} also shows the data acquired in a single crystal.  The data clearly are reproducible across different samples, and our observations are independent of whether the sample has been subjected to strain in order to form the powder.
We note that other groups have published similar data,
albeit with less precision, but have not speculated on the origin of this suppression. \cite{kohoriURu2Si2,Baek2010a}


In order to quantify the observed pseudogap, it is important to clarify the origin of the spin-lattice-relaxation in the paramagnetic state. Inelastic neutron scattering experiments have shown the importance of spin fluctuations in URu$_2$Si$_2$ and their possible connection to the HO phase transition. \cite{BroholmURS, WiebeURS}
Because of strong spin-orbit coupling neither spin nor orbital quantum numbers are conserved and instead fluctuations of the magnetic moment aligned along the $c$ axis should be considered as pseudospin fluctuations, \cite{chandraURS,zhitomirsky,Elgazzar2008,Das2012,Chandra2013,RiseburoughURS} where these fluctuations are based on the spin-orbit interaction of the $j$-$j$ coupling scheme for the $j=5/2$ sextet. \cite{HottaPRB2003}
Furthermore, we will assume that these fluctuations are the signature of a quantum phase transition (QPT), which is related to the HO transition and long-range antiferromagnetic (AFM) order found at pressure above $0.6$ GPa. \cite{Amitsuka2007214} Although this QPT is complicated by a first-order phase transition under pressure from the HO to AFM phase, the normal state to HO transition remains second order. Further corroborating the QPT scenario is the strong coupling between the HO and AFM in Rh-doped studies, which implies that there is a nearby instability to AFM at ambient pressure. \cite{Baek2010a} Such a scenario of dominant 2D fluctuations is supported by the noticeable anisotropy of magnetic correlation lengths $\xi_a/\xi_c \sim 4$, deduced from scans around neutron scattering peaks (1,0,0) and (1,0,2). \cite{BroholmURS} Similar two-dimensional (2D) correlations were reported for the heavy fermion compounds CeCu$_{6-x}$Au$_x$ and YbRh$_2$Si$_2$. \cite{StockertURS1998,Gegenwart2008}

\begin{figure}
\includegraphics[width=\linewidth]{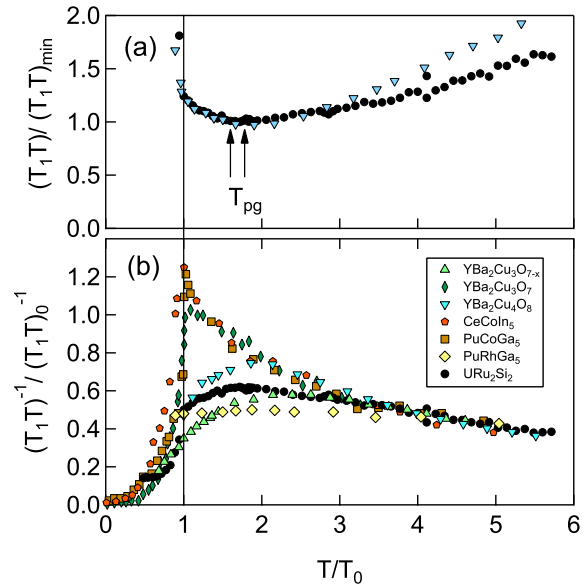}
\caption{(Color Online) (a) ($T_1T$)/($T_1T$)$_{min}$ as function of $T/T_0$ for URu$_2$Si$_2$  and YBa$_2$Cu$_4$O$_8$. The arrows indicate the onset of the pseudogap in both materials. (b) The scaling behavior of ($T_1T$)$^{-1}$/($T_1T$)$_0^{-1}$ as function of $T/T_0$ for URu$_2$Si$_2$ (black circles) as compared to other materials with AFM fluctuations. }
\label{fig:scaling}
\end{figure}

Given the nature of 2D AFM pseudospin fluctuations, we argue that the longitudinal fluctuations of the $z$ component of the total angular momentum ($\Delta j_z =\pm 1$),
seen in the four-point correlation function describing the $(T_1T)^{-1}$ data of Fig.~\ref{fig:T1Tinv},
can be captured by the two-component spin-fermion model. \cite{bangT1pumga5} In Bang's model, critical spin modes originate from localized $f$-electron spins coupled to conduction electrons. Since it successfully explained the spin-lattice-relaxation rate in Pu$M$Ga$_5$ ($M$=Rh, Co) compounds (shown in Fig.~\ref{fig:scaling}), we expect the same to apply to URu$_2$Si$_2$:
\begin{equation}
(T_1T)^{-1} \sim N_J(T) \times \xi(T)^z .
\label{eqn:T1formula}
\end{equation}
Here $N_J(T)\sim N^2(E_F)$ is the normalized joint density of states (JDOS) at the Fermi energy $E_F$, which accounts for any $T$-dependent suppression due to a pseudogap, and $\xi(T)$ is the magnetic correlation length of dynamic fluctuations with dynamic exponent $z$. The slowing down of fluctuations is given by the relaxation time $\tau \sim \xi^z$. For commensurate 2D AFM Heisenberg fluctuations near a QPT, one expects to find $z=1$.
In this case, the magnetic correlation length shows a universal scaling behavior when crossing from the quantum critical to quantum disordered state and is approximately given by: \cite{CHN}
\begin{equation}
\xi(T)^{-1} \sim {\Delta+T\exp(-4\Delta/T)},
\label{eqn:xi}
\end{equation}
where $\Delta \sim T_{QPT} $ is the pseudospin gap in the excitation spectrum of the quantum critical state above the crossover.  Note that at exactly the critical point ($\Delta=0$), the correlation length obeys
$\xi^{-1} \sim T$ as $T\to 0$, characteristic of QPT phenomena.


We extract the hidden order and pseudogap suppressed JDOS $N_J(T)$ from the spin-lattice-relaxation rate. First we determine the spin gap from the fluctuation region above the onset of the pseudogap. Above 30 K, we fit $(T_1 T)^{-1}$ to the expression in Eqs.~(\ref{eqn:T1formula}) and (\ref{eqn:xi}) with dynamic exponent $z=1$:
\begin{eqnarray}
1/T_1 T = N_J(T)
\frac{A}{\Delta+T\exp(-4\Delta/T)} + B ,
\label{eqn:T1fit}
\end{eqnarray}
with positive coefficients $A$ and $B$, where $B$ is {a high temperature Korringa relaxation term} that accounts for conduction bands not participating in the fluctuations, {and assume that $N_J=1$}.
We find that the data in Fig.~\ref{fig:T1Tinv} are best described by model parameters
$\Delta \approx 1.9 \, T_{HO}= 30.1$ K,  $A=1.41\, {\rm s^{-1}}$, $B=9.03\times 10^{-3} \, {\rm K^{-1} s^{-1}}$ and $N_J(T)=1$ for $T>30$ K
{using the least-absolute deviation (LAD) method, rather than the least-square (LSQ) method,}
because it is less sensitive to scatter in the data, as shown in Fig.~\ref{fig:T1Tinv}.
{While both LAD and LSQ give good overall fits over the entire temperature range, the LSQ fit yields a residual JDOS that is too small (nearly zero) to be consistent with specific heat measurements of a heavy mass and heavy-fermion superconductivity below 1.5 K. The primary difference  between both fits is in the value of the $B$ term. Furthermore, restricting the fitting window to higher temperatures leads to unphysically small Curie-Weiss temperatures  that clearly contradict values obtained from extrapolation at higher temperatures, i.e., $100 - 300$ K. \cite{PalstraURSdiscovery} }
A comparison with Curie-Weiss-like local moment fluctuations, $\xi\sim 1/(T+T_{CW})$ and $T_{CW}\approx 90$ K, gives the worst fit and an unreasonably large residual JDOS in the HO state.
{Note that $T_{CW}$ is similar to $T_{coh}$ as measured by the Knight shift, where local moment behavior emerges above $T_{coh}$. \cite{ShirerPNAS2012} }
{Next, we apply the fluctuation theory over the entire temperature range and divide out the experimental data to yield $N_J(T)$ by using Eq.~(3) as shown in Fig.~\ref{fig:T1Tinv}(b).}

Clearly, the suppression of $(T_1 T)^{-1}$ between
{$T_{pg}\sim 30$ K and $T_{HO}\sim 16$ K}
necessitates the existence of a pseudogap, so that $N_J(T)<1$ for $T<T_{pg}$, which reflects a suppression of the modes of the pseudospin fluctuations.
{Between $T_{pg}$ and $T_{HO}$ the JDOS is suppressed 25\% and
continues to be suppressed even more below $T_{HO}$, however, it remains finite
($\sim 6$\%) at 8.5 K, see the LAD curve in Fig.~2(b),}
in agreement with specific heat and STM measurements. \cite{DavisURSnature,YazdaniURS} The anomalous $T$-dependence of $T_1$ in the paramagnetic state also shows up in many other correlated electron systems as exemplified in Fig.~\ref{fig:scaling}. Optical measurements of URu$_2$Si$_2$ have also observed non-Fermi liquid behavior, as one might expect for a QPT. \cite{TimuskURS2012} In the high temperature superconducting cuprates, the pseudogap is manifested in the spin-lattice-relaxation rate data as a deviation from linearity in a plot of $T_1T$ versus $T$. \cite{Corey1996} In fact, the qualitative behavior of $T_1T$ in URu$_2$Si$_2$ is quite similar to that of YBa$_2$Cu$_4$O$_8$, as seen in Fig.~\ref{fig:scaling}(a). In this case, the temperature axis is scaled by $T_0$, where $T_0$ is either $T_{HO}$ or $T_c$,
{the latter is the}
superconducting transition temperature. For both materials, $T_1T$ exhibits a clear upturn at $\sim 1.8T_0$. For $T>T_{pg}$, the qualitative similarity is even more striking. Fig.~\ref{fig:scaling}(b) shows the inverse, $(T_1T)^{-1}$, versus $T/T_0$ for several high-temperature superconductors, heavy-fermion superconductors, and URu$_2$Si$_2$.
{Although we cannot determine the origin if the observed pseudogap, we do note that} when $(T_1T)^{-1}$ is normalized
to remove the effects of different hyperfine couplings
[$(T_1T)^{-1}_0 = 1/2$ at $T=4T_0$, where $T_0$ is the ordering temperature],
the data scale with one another in the normal (paramagnetic) states for $T> 3T_0$,
suggesting a connection between the nature of the ordering and the AFM pseudospin fluctuations driving the relaxation rate. \cite{Curro2005}
Since the effects of 2D AFM spin fluctuations play a crucial role in a large class of materials above the superconducting transition, and possibly in the mediation of superconductivity, it is quite likely that the long-range HO state evolves out of pseudospin fluctuations as well.
Recent theoretical scenarios falling into this general class of fluctuation mediated HO state are the spin-orbit-density and hybridization wave, \cite{Das2012,RiseburoughURS,Chandra2013,Elgazzar2008,DubiBalatskyPRL2011} and pseudospin-density wave due to crystal-field split ground states. \cite{zhitomirsky}

In conclusion, we have demonstrated the existence of a hidden order pseudospin gap  $\Delta \approx
1.9\, T_{HO}$ = 30.1 K  that is consistent with other probes,
extracted the JDOS through a fitting of the NMR relaxation time data,
{and} compared this {behavior} to other materials with known pseudogap behavior. It is our hope that other {experiments} will investigate this region and the connection between the hidden order and antiferromagnetic state and the proposed pseudogap behavior.

We thank T. Das, P. Coleman, P. Riseborough,  J. Mydosh, and T. Durakiewicz for stimulating discussions. Work at UC Davis and LANL was supported by the UC Lab Research Fee Program and the NNSA under the Stewardship Science Academic Alliances program through U.S. DOE Research Grant
No.\ DE-FG52-09NA29464.  JTH and AVB acknowledge support for work performed, in part, by the Center for Integrated Nanotechnologies, an Office of Science User Facility operated for the U.S. DOE Office of Science by LANL (Contract No.\ DE-AC52-06NA25396) and AVB acknowledges support by Nordita. MJ gratefully acknowledges financial support by the Alexander von Humboldt foundation. Research at UCSD was supported by the U. S. Department of Energy under Grant No.\ DE-FG02-04ER46105.


\bibliography{URu2Si2_NMRT1_r3}

\end{document}